\documentclass[pra,twocolumn,superscriptaddress]{revtex4-1}%
\usepackage{amsfonts}
\usepackage[colorlinks=true,linkcolor=blue,citecolor=red,plainpages=false,pdfpagelabels]%
{hyperref}
\usepackage{amsmath}
\usepackage{amssymb}
\usepackage{graphicx}%
\providecommand{\U}[1]{\protect\rule{.1in}{.1in}}
\newtheorem{theorem}{Theorem}

\newtheorem{definition}[theorem]{Definition}

\newtheorem{lemma}[theorem]{Lemma}

\newtheorem{property}{Property}[section]

\newcommand{\ket}[1]{| #1 \rangle}
\newcommand{\bra}[1]{\langle #1 |}

\def\U{\mathrm{U}}

\def\C{\mathcal{C}}
\def\D{\mathcal{D}}

\def\G{\mathcal{G}}
\def\H{\mathcal{H}}
\def\U{\mathcal{U}}

\def\N{\mathcal{N}}

\def\S{\mathcal{S}}

\def\X{\mathcal{X}}

\newcommand{\tr}{\operatorname{Tr}}

\begin{document}
\preprint{ }
\title{Conditional quantum one-time pad}
\author{Kunal Sharma}
\affiliation{Hearne Institute for Theoretical Physics, Department of Physics and Astronomy, and Center for Computation and Technology,
Louisiana State University, Baton Rouge, Louisiana 70803, USA}
\author{Eyuri Wakakuwa}
\affiliation{Graduate School of Informatics and Engineering, University of
Electro-Communications, 1-5-1 Chofugaoka, Chofu-shi, Tokyo, 182-8585, Japan}
\author{Mark M. Wilde}
\affiliation{Hearne Institute for Theoretical Physics, Department of Physics and Astronomy, and Center for Computation and Technology,
Louisiana State University, Baton Rouge, Louisiana 70803, USA}
\keywords{conditional quantum one-time pad, conditional quantum mutual information}
\pacs{}

\begin{abstract}
Suppose that Alice and Bob are located in distant laboratories, which are
connected by an ideal quantum channel. Suppose further that they share many
copies of a quantum state $\rho_{ABE}$, such that Alice possesses the $A$
systems and Bob the $BE$ systems. In our model, there is an identifiable part
of Bob's laboratory that is insecure:\ a third party named Eve has infiltrated
Bob's laboratory and gained control of the $E$ systems. Alice, knowing this,
would like to use their shared state and the ideal quantum channel to communicate
a message in such a way that Bob, who has access to the whole of his
laboratory ($BE$ systems), can decode it, while Eve, who has access only to a
sector of Bob's laboratory ($E$ systems)\ and the ideal quantum channel
connecting Alice to Bob, cannot learn anything about Alice's transmitted
message. We call this task the conditional one-time pad, and in this paper, we
prove that the optimal rate of secret communication for this task is equal to
the conditional quantum mutual information $I(A;B|E)$\ of their shared state.
We thus give the conditional quantum mutual information an operational meaning
that is different from those given in prior works, via state redistribution,
conditional erasure, or state deconstruction. We also generalize the model and
method in several ways, one of which is a secret-sharing task,
i.e., the case in which Alice's message should be secure from someone
possessing only the $AB$ or $AE$ systems but should be decodable by someone
possessing all systems $A$, $B$, and $E$.

\end{abstract}
\volumeyear{ }
\volumenumber{ }
\issuenumber{ }
\eid{ }
\date{\today}
\startpage{1}
\endpage{10}
\maketitle

\textit{Introduction}---This paper  shows that the
optimal rate of a communication task, which we call the conditional one-time
pad, is equal to a fundamental information quantity called the conditional
quantum mutual information. To prove this statement, we operate in the regime
of quantum Shannon theory \cite{H06book,H12,W15book}, supposing that Alice and Bob possess a large
number $n$ of copies of a quantum state $\rho_{ABE}$. We suppose that one
party Alice has access to all of the $A$ systems, and another party Bob has
access to all of the $BE$ systems. We suppose that Bob's laboratory is divided
into two parts, one of which is secure (the $B$ part)\ and the other which is
insecure (the $E$ part) and accessible to an eavesdropper Eve. We also suppose
that Alice and Bob are connected by an ideal quantum channel, but the
eavesdropper Eve can observe any quantum system that is transmitted over the
ideal channel if she so desires. The goal of a conditional quantum one-time
pad protocol is for Alice to encode a message $m$ into her $A$ systems, in
such a way that if she sends her $A$\ systems over the ideal quantum channel, then

\begin{enumerate}
\item Bob can decode the message $m$ reliably by performing a measurement on
all of the $ABE$\ systems, while

\item an eavesdropper possessing the $AE$ systems has essentially no chance of
determining the message $m$ if she tried to figure it out.
\end{enumerate}

\noindent We prove that the optimal asymptotic rate at which this task can be
accomplished is equal to the conditional quantum mutual information of the
state $\rho_{ABE}$, defined as%
\begin{equation}
I(A;B|E)_{\rho}\equiv I(A;BE)_{\rho}-I(A;E)_{\rho},\label{eq:CQMI-def}%
\end{equation}
where the quantum mutual information of a state $\sigma_{FG}$ is defined as
$I(F;G)_{\sigma}\equiv H(F)_{\sigma}+H(G)_{\sigma}-H(FG)_{\sigma}$, with
$H(F)_{\sigma}\equiv-\operatorname{Tr}\{\sigma_{F}\log_{2}\sigma_{F}\}$
denoting the quantum entropy of the reduced state $\sigma_F$. 

Our main result thus gives an operational meaning to the conditional quantum
mutual information (CQMI)\ that is conceptually different from those appearing
in prior works \cite{DY08,YD09,BBMW16a,BBMW16}. CQMI has previously been interpreted
as the optimal rate of quantum communication from a sender to a receiver to
accomplish the task of state redistribution \cite{DY08,YD09}, in which the
goal is for a sender to transmit one of her systems to a receiver who
possesses a system correlated with the systems of the sender. CQMI\ has also
been interpreted as the optimal rate of noise needed to accomplish the task of
conditional erasure or state deconstruction \cite{BBMW16a,BBMW16}, in which
(briefly)\ the goal is to apply noise to the $AE$ systems of $\rho
_{ABE}^{\otimes n}$ such that the resulting $A$ systems are locally
recoverable from the $E$ systems alone while the marginal state $\rho
_{BE}^{\otimes n}$ is negligibly disturbed. Recently, the dynamic counterpart of CQMI has been interpreted as the optimal rate of entanglement-assisted private communication over quantum broadcast channels \cite{QSW18}, which is inspired by the conditional one-time pad protocol presented in this work. 

The conditional mutual information is an information quantity that plays a
central role in quantum information theory. The fact that it is non-negative
for any quantum state is non-trivial and known as the strong subadditivity of
quantum entropy \cite{LR73,PhysRevLett.30.434}. The strong subadditivity
inequality is at the core of nearly every coding theorem in quantum
information theory (see, e.g., \cite{H12,H06book,W15book}). The CQMI\ is also the
information quantity underlying an entanglement measure called squashed
entanglement \cite{Christandl2003}, a quantum correlation measure called
quantum discord \cite{Z00,zurek01}\ (as shown in \cite{P12}), and a steering
quantifier called intrinsic steerability \cite{KWW16}. The CQMI is also a witness of Markovianity in the sense that if $I(A;B \vert E)$ is small, then the correlations between systems $A$ and $B$ are mediated by the system $E$ via a recovery channel from $E$ to $AE$ \cite{FR15}. Moreover, the CQMI of three regions with a non-trivial topology leads to the topological entanglement entropy of the system, which essentially characterizes irreducible many-body correlation \cite{KP06, LW06, K12}. The CQMI is thus an important information quantity to study quantum correlations in condensed matter systems (see, e.g., \cite{ZCZW15}). Furthermore, in the context of thermodynamics, the CQMI has been used to establish that the free fermion non-equilibrium steady state is an approximate quantum Markov chain \cite{MFMTS16}. The CQMI also plays an important role in high energy physics \cite{CLMS15, Ding2016, PEW17}.

The basic intuition for the achievability of the conditional mutual
information for the conditional one-time pad task is obtained by inspecting
the expansion in \eqref{eq:CQMI-def}\ and is as follows: the authors of
\cite{SW06}\ showed that the quantum mutual information of a bipartite state
is equal to the optimal rate of a task they called the (unconditional) quantum
one-time pad. In our setting, the result of \cite{SW06} implies that Alice can
communicate a message secure against an eavesdropper, who can observe only the
$A$ systems, such that Bob, in possession of the $BE$ systems, can decode it
reliably, as long as the number of messages is $\approx nI(A;BE)_{\rho}$ bits.
Here, we show that the message of Alice can be secured against an eavesdropper
having access to both the $A$ and $E$ systems if Alice sacrifices $\approx
nI(A;E)_{\rho}$ bits of the message, such that the total number of bits of the
message is $\approx nI(A;BE)_{\rho}-nI(A;E)_{\rho}=nI(A;B|E)_{\rho}$, where we
have employed \eqref{eq:CQMI-def}. The main idea for a code construction to
accomplish the above task is the same as that for the classical wiretap
channel \cite{W75}, which has been extended in a certain way to the quantum
case in \cite{ieee2005dev,1050633}. To prove the achievability part of the
main result of our paper, we use a coding technique developed in
\cite[Section~III-A]{itit2008hsieh} and which was rediscovered shortly
thereafter in \cite{SW06} and later used in \cite{DTW14}. We also employ tools
known as the quantum packing and covering lemmas (see,
e.g., \cite{W15book}). To establish optimality of the CQMI for the
conditional one-time pad task, we employ entropy inequalities. We
note that the aforementioned methods also lead to a proof of the main result
of \cite{BO12}, which concerns a kind of quantum one-time pad protocol
different from that developed in \cite{SW06}\ or the present paper.

A modification of the coding structure for the conditional one-time pad
protocol allows us to establish that the following 
information quantity%
\begin{equation}
I(A;BE)_{\rho}-\max\{I(A;B)_{\rho},I(A;E)_{\rho}\}
\label{eq:info-scramble-rate}
\end{equation}
of a tripartite state $\rho_{ABE}$ is an optimal achievable rate for a particular
secret-sharing task that we call \textit{information scrambling}. In this
modified task, we suppose that Alice, Bob, and Eve are three distinct parties.
Alice's laboratory is distant from Bob and Eve's, but we imagine that Bob and
Eve's laboratories are close together, and an ideal quantum channel connects
Alice's laboratory to Bob and Eve's. The goal of the information scrambling
task is for Alice to communicate a message in such a way that it can be
decoded only by someone who possesses all three $ABE\ $systems. If someone
possesses only the $AB$ systems or only the $AE$ systems, then such a person
can figure out essentially nothing about the encoded message.

Our finding here shows that the quantity in \eqref{eq:info-scramble-rate} is an optimal achievable rate for information scrambling, such that
the message is encoded in the non-local degrees of freedom of $\rho
_{ABE}^{\otimes n}$ and cannot be decoded exclusively from the local degrees
of freedom, which in this case are constituted by systems $AB$ or systems $AE$.

The rest of our paper proceeds as follows. We first formally define the
conditional one-time pad task.  We then
sketch a proof for the achievability part of our result. We finally discuss variations of the main task, such as the
information scrambling task mentioned above and more general tasks, and then
we conclude with a brief summary.

The supplementary material provides a detailed proof of the achievability part of our main result.
It also establishes the optimality part of
our main result:\ that Alice cannot communicate at a rate higher than the
conditional mutual information $I(A;B|E)$ while still satisfying the joint
demands of reliable decoding for Bob (who gets the $ABE$ systems) and security
against an eavesdropper who has access to the $AE$ systems. The optimality proof is based on entropy inequalities and identities.

\textit{Conditional quantum one-time pad}---We use notation and concepts
standard in quantum information theory and point the reader to \cite{W15book}%
\ for further background. Let $n,M\in\mathbb{N}$ and let $\varepsilon
,\delta\in[0,1]$. An $(n,M,\varepsilon,\delta)$ conditional one-time pad
protocol begins with Alice and Bob sharing $n$ copies of the state $\rho
_{ABE}$, so that their state is $\rho_{ABE}^{\otimes n}$. As mentioned
previously, Bob has access to the $BE$ systems, but we consider the $E$
systems to be insecure and jointly accessible by an eavesdropper. Alice and
Bob are connected by an ideal quantum channel, which Eve has access to as well (later we argue that it suffices
for Alice and Bob to use only $\approx nH(A)_{\rho}$ ideal qubit channels, but for now we
suppose that the ideal quantum channel can transmit as many qubits as
desired). At the beginning of the protocol, Alice picks a message
$m\in\{1,\ldots,M\}$ and applies an encoding channel $\mathcal{E}%
_{A^{n}\rightarrow A^{\prime}}^{m}$ to the $A^{n}$ systems of $\rho
_{ABE}^{\otimes n}$, leading to the state $\omega_{A^{\prime}B^{n}E^{n}}%
^{m}\equiv\mathcal{E}_{A^{n}\rightarrow A^{\prime}}^{m}(\rho_{ABE}^{\otimes
n})$. She transmits the system $A^{\prime}$ of $\omega_{A^{\prime}B^{n}E^{n}%
}^{m}$ over the ideal quantum channel. Bob applies a decoding positive
operator-valued measure $\{\Lambda_{A^{\prime}B^{n}E^{n}}^{m}\}_{m}$ to the
systems $A^{\prime}B^{n}E^{n}$ of $\omega_{A^{\prime}B^{n}E^{n}}^{m}$ in order
to figure out which message was transmitted. The protocol is $\varepsilon
$-reliable if Bob can determine the message $m$ with probability not smaller
than $1-\varepsilon$:%
\begin{equation}
\forall m:\operatorname{Tr}\{\Lambda_{A^{\prime}B^{n}E^{n}}^{m}\omega
_{A^{\prime}B^{n}E^{n}}^{m}\}\geq1-\varepsilon.\label{eq:reliability-cond}%
\end{equation}
The protocol is $\delta$-secure if the reduced state $\omega_{A^{\prime}E^{n}%
}^{m}$\ on systems $A^{\prime}E^{n}$ is nearly indistinguishable from a
constant state $\sigma_{A^{\prime}E^{n}}$ independent of the message $m$:%
\begin{equation}
\forall m:\frac{1}{2}\left\Vert \omega_{A^{\prime}E^{n}}^{m}-\sigma
_{A^{\prime}E^{n}}\right\Vert _{1}\leq\delta,\label{eq:security-cond}%
\end{equation}
where we have employed the normalized trace distance.

We say that a rate $R$ is achievable for the conditional quantum one-time pad
if for all $\varepsilon,\delta\in(0,1)$, $\gamma>0$, and sufficiently large
$n$, there exists an $(n,2^{n\left[  R-\gamma\right]  },\varepsilon,\delta)$
conditional one-time pad protocol of the above form. The conditional one-time
pad capacity of a state $\rho_{ABE}$ is equal to the supremum of all
achievable rates.

\textit{Achievability of CQMI for conditional one-time pad}---Here we mostly
sketch an argument that the CQMI $I(A;B|E)_{\rho}$ is a lower bound on the
conditional one-time pad capacity of $\rho_{ABE}$, while the supplementary material contains a detailed proof. First, consider the reduced state $\rho_{A}$ and a
spectral decomposition for it as $\rho_{A}=\sum_{x}p_{X}(x)|x\rangle\langle
x|_{A}$, where $p_{X}$ is a probability distribution and $\{|x\rangle
_{A}\}_{x}$ is an orthonormal basis. Let $|\phi\rangle_{AR}=\sum_{x}%
\sqrt{p_{X}(x)}|x\rangle_{A}|x\rangle_{R}$ be a purification of $\rho_{A}$.
Let $|\psi\rangle_{ABEF}$ denote a purification of $\rho_{ABE}$, with $F$
playing the role of a purifying system. Since all purifications are related by
an isometry acting on the purifying system, there exists an isometry
$U_{R\rightarrow BEF}$ such that $U_{R\rightarrow BEF}|\phi\rangle_{AR}%
=|\psi\rangle_{ABEF}$. Applying the isometry $U_{R\rightarrow BEF}$ followed
by a partial trace over $F$ can be thought of as a channel $\mathcal{N}%
_{R\rightarrow BE}$ that realizes the state $\rho_{ABE}$ as $\mathcal{N}%
_{R\rightarrow BE}(\phi_{AR})=\rho_{ABE}$. Similarly, if we apply the isometry
$U_{R\rightarrow BEF}$ and trace over $FB$, then this is a channel
$\mathcal{M}_{R\rightarrow E}$ that realizes the reduced state $\rho_{AE}$ as
$\mathcal{M}_{R\rightarrow E}(\phi_{AR})=\rho_{AE}$.

If we take $n$ copies of $\rho_{ABE}$, then the state $\rho_{ABE}^{\otimes n}$
can be thought of as the following state $\mathcal{N}_{R\rightarrow
BE}^{\otimes n}(\phi_{AR}^{\otimes n})$. The pure state $|\phi\rangle
_{AR}^{\otimes n}$ admits an information-theoretic type decomposition of the
following form:
$
|\phi\rangle_{AR}^{\otimes n}=\sum_{t}\sqrt{p(t)}|\Phi_{t}\rangle_{A^{n}R^{n}%
},
$
where the label $t$ indicates a type class and $|\Phi_{t}\rangle_{A^{n}R^{n}}$
is a maximally entangled state of Schmidt rank $d_{t}$ with support on the
type class subspace labeled by $t$. We can then consider forming encoding
unitaries out of the generalized Pauli shift and phase-shift operators
\begin{equation}
V_{A^{n}}(x_{t},z_{t})=X_{A^{n}}(x_{t})Z_{A^{n}}(z_{t}),
\end{equation}
which act on a
given type class subspace $t$ and where $x_{t},z_{t}\in\{0,\ldots,d_{t}-1\}$.
The overall encoding unitary allows for an additional phase $(-1)^{b_{t}}$ for
$b_{t}\in\{0,1\}$ and has the form
\begin{equation}
U_{A^{n}}(s)=\bigoplus\limits_{t}%
(-1)^{b_{t}}V_{A^{n}}(x_{t},z_{t}),
\end{equation}
where $s$ is a vector $[(b_{t}%
,x_{t},z_{t})]_{t}$.

The coding scheme is based on random coding, as is usually the case in quantum
Shannon theory, and works as follows. Let $M,K\in\mathbb{N}$. Alice has a
message variable $m\in\{1,\ldots,M\}$ and a local key variable $k\in
\{1,\ldots,K\}$. For each pair $(m,k)$, Alice picks a vector $s$, of the form
described previously, uniformly at random and labels it as $s(m,k)$. The set
$\mathcal{C}=\{s(m,k)\}_{m,k}$ constitutes the code, and observe that it is
initially selected randomly. If Alice wishes to send message $m$, then she
picks $k$ uniformly at random from $k\in\{1,\ldots,K\}$, applies the encoding
unitary $U_{A^{n}}(s(m,k))$ to the state $\rho_{ABE}^{\otimes n}$ and sends
the $A^{n}$ systems to Bob. Bob's goal is to decode both the message variable
$m$ and the local key variable $k$. Based on the packing lemma, it follows that if $\log_{2}MK\approx nI(A;BE)_{\rho}$, then there
is a decoding measurement $\{\Lambda_{A^{n}B^{n}E^{n}}^{m,k}\}$ for Bob,
constructed from typical projectors and corresponding to a particular selected
code $\mathcal{C}$, such that
\begin{multline}
\mathbb{E}_{\mathcal{C}}\Bigg\{\frac{1}{MK}\sum_{m,k}\operatorname{Tr}%
\{\Lambda_{A^{n}B^{n}E^{n}}^{m,k}U_{A^{n}}(S(m,k))\rho_{ABE}^{\otimes n}%
\times\\
U_{A^{n}}^{\dag}(S(m,k))\}\Bigg\}\geq1-\varepsilon,
\end{multline}
for all $\varepsilon\in(0,1)$ and sufficiently large $n$, and where the
expectation is with respect to the random choice of code $\mathcal{C}$. On the
other hand, from the perspective of someone who does not know the choice of
$k$ and who does not have access to the systems $B^{n}$, the state has the
following form:%
\begin{equation}
\tau_{A^{n}E^{n}}^{m}\equiv\frac{1}{K}\sum_{k=1}^{K}U_{A^{n}}(s(m,k))\rho
_{AE}^{\otimes n}U_{A^{n}}^{\dag}(s(m,k)).
\end{equation}
The quantum covering lemma and the
properties of typical projectors guarantee that%
\begin{multline}
\Pr_{\mathcal{C}}\{\left\Vert \tau_{A^{n}E^{n}}^{m}-\overline{\tau}_{A^{n}E^{n}}\right\Vert
_{1}\leq\delta+4\sqrt{\delta}+24\sqrt[4]{\delta}\}\\
\geq1-2D\exp\left(  -\frac{\delta^{3}K2^{-n[I(A;E)_{\rho}+\delta']}}{4}\right)  ,
\end{multline}
where $D$ is a parameter that is no more than exponential in $n$, $\delta' >0$ is a small constant, and 
\begin{equation}
\overline{\tau}_{A^{n}E^{n}}\equiv\mathbb{E}_{S}\{  U_{A^{n}}(S)\rho
_{AE}^{\otimes n}U_{A^{n}}^{\dag}(S)\}  .
\end{equation}
Thus, as long as we pick $\log_{2}K\approx nI(A;E)_{\rho}$, then there is an
extremely good chance that the state $\tau_{A^{n}E^{n}}^{m}$ will be nearly
indistinguishable from the average state $\overline{\tau}_{A^{n}E^{n}}$.
Now, we can define the
event $E_{0}$ to be the event that Bob's measurement decodes with high average success
probability and the event $E_{m}$ to be the event that $\left\Vert \tau
_{A^{n}E^{n}}^{m}-\overline{\tau}_{A^{n}E^{n}}\right\Vert _{1}$ is small. The
union bound of probability theory then guarantees that there is a non-zero probability for there to be a
code $\{s(m,k)\}_{m,k}$ such that the average success probability of
Bob's decoder is arbitrarily high and $\left\Vert \tau_{A^{n}E^{n}}%
^{m}-\overline{\tau}_{A^{n}E^{n}}\right\Vert _{1}$ is arbitrarily small for
all $m$, with these statements holding for sufficiently large $n$. So this means that such a code $\{s(m,k)\}_{m,k}$ exists. A final
\textquotedblleft expurgation\textquotedblright\ argument guarantees that Bob can decode each $m$ and $k$
with arbitrarily high probability and that $\left\Vert \tau_{A^{n}E^{n}}%
^{m}-\overline{\tau}_{A^{n}E^{n}}\right\Vert _{1}$ is arbitrarily small for
all $m$. Therefore, the number of bits that Alice can communicate securely is thus
\begin{align}
\log_{2}M & =\log_{2}MK-\log_{2}K \notag \\
   & \approx nI(A;BE)_{\rho}-nI(A;E)_{\rho
} \notag \\
& =nI(A;B|E)_{\rho},
\end{align}
 so that $I(A;B|E)_{\rho}$ is an achievable rate. This
concludes the achievability proof sketch. As indicated previously, the optimality proof is given in the supplementary material.

We note that it actually suffices to use $\approx nH(A)_{\rho}$ noiseless
qubit channels for the communication of the $A$ systems, rather than
$n\log\left\vert A\right\vert $ noiseless qubit channels. This is because
Alice can perform Schumacher compression \cite{PhysRevA.51.2738} of her $A^{n}$ systems before
transmitting them, and the structure of the encoding unitaries is such that
this can be done regardless of which message is being transmitted (see the
discussion at the end of \cite[Section~22.3]{W15book}). The Schumacher compression causes a negligible disturbance to each of the states that is transmitted.

\textit{Conditional one-time pad of a quantum message}---We note that it is
possible to define a conditional quantum one-time pad of a quantum message, in
which the goal is to transmit one share $\hat{M}$\ of a quantum state
$|\varphi\rangle_{M^{\prime\prime}\hat{M}}$ securely in such a way that Bob,
possessing systems $A^{\prime}B^{n}E^{n}$, can decode the quantum message in
$\hat{M}$, while someone possessing the systems $A^{\prime}E^{n}$ cannot learn
anything about the quantum system $\hat{M}$. Our result here is that
$I(A;B|E)_{\rho}/2$ is the optimal rate for this task of conditional one-time
pad of a quantum message. The optimality proof is nearly identical to the optimality
proof given previously, except that we start with the assumption that the initial
state $|\varphi\rangle_{M^{\prime\prime}\hat{M}}$ is a maximally entangled
state $|\Phi\rangle_{M^{\prime\prime}\hat{M}}$, such that the quantum
information in system $\hat{M}$ can be decoded well. Then, the proof starts
with the condition that $\log_{2}M=I(M^{\prime\prime};\hat{M})_{\Phi}/2$ and
proceeds identically from there. For the achievability part, we perform a
coherent version of the above protocol, as reviewed in \cite[Section~22.4]%
{W15book}, and we find that it generates coherent bits \cite{prl2004harrow}%
,\ which are secure from someone possessing the $A^{n}E^{n}$ systems, at a
rate equal to $I(A;B|E)_{\rho}$. By the coherent communication identity from
\cite{prl2004harrow}, it follows that qubits can be transmitted securely at a
rate equal to $I(A;B|E)_{\rho}/2$.

\textit{Generalizations}---We note that the coding scheme outlined above in
the achievability proof can be generalized in several interesting ways.
Suppose that Alice shares a state with \textquotedblleft many
Bobs\textquotedblright, i.e., one of the form $\rho_{AB_{1}\cdots B_{\ell}}$
for some positive integer $\ell\geq2$. Then Alice might wish to encode a
message $m$ in her $A$ systems of $\rho_{AB_{1}\cdots B_{\ell}}^{\otimes n}$
in such a way that only someone possessing all of the systems $AB_{1}\cdots
B_{\ell}$ would be able to decode it, but someone possessing system $A$ and
some subset $\mathcal{B}_{i}\in\{B_{1},\ldots,B_{\ell}\}$ would not be able
to determine anything about the message $m$. Alice might wish to protect the message against several different subsets
$\mathcal{B}_{i}$, for $i\in\{1,\ldots,p\}$, as in secret sharing. Then we could structure a coding scheme similar to our achievability proof to have a message variable $m\in\{1,\ldots,M\}$ and a local key variable $k \in \{1, \ldots, K \}$, such that
\begin{align}
\log_{2}MK & \approx nI(A;B_{1}\cdots B_{\ell}),\\
\log_{2}K & \approx n[\max \{ I(A;\mathcal{B}_{1}), \dots, I(A;\mathcal{B}_{p}) \}].
\end{align}
  Given that
  \begin{equation}
  I(A;B_{1} \cdots B_{\ell})_{\rho} -  \max \{ I(A;\mathcal{B}_{1})_{\rho}, \dots, I(A;\mathcal{B}_{p})_{\rho} \}
  \end{equation}
   is always non-negative, the coding scheme guarantees that this information difference is an achievable rate that accomplishes the desired task. We note that the secret-sharing task discussed above is different from the previously considered protocols in \cite{HBB99, SS18}, and references therein.

A particular case of interest is the scenario mentioned earlier in the paper
and which we called information scrambling. There, Alice, Bob, and Eve share a state $\rho_{ABE}$,
and the goal is for Alice to encode a message in the $A$ system such that
someone possessing the $ABE$ systems can decode it, but someone possessing the
$AB$ systems or the $AE$ systems cannot
determine anything about the message $m$ (i.e., the message $m$ has been
scrambled in the nonlocal degrees of freedom of the state $\rho_{ABE}$ and is
not available in $\rho_{AB}$ or $\rho_{AE}$). According to the above
reasoning, an achievable rate for this task is the information quantity $I(A;BE)_{\rho}- \max\{I(A;B)_{\rho
},I(A;E)_{\rho}\}$. This rate is also optimal.

We note also that our methods give a concrete and transparent approach to
prove the results of \cite{BO12}, as discussed in the supplementary material. In particular, we have established an
information-theoretic converse of that result using entropy identities and
inequalities along the lines presented previously, and the achievability part of
that result can be accomplished by using the encoding unitaries discussed
earlier, along with the quantum packing and covering lemmas.

Our operational interpretation of the conditional mutual information also
leads to an interesting operational interpretation of the squashed
entanglement of a bipartite state $\rho_{AB}$: we can consider squashed
entanglement to be the optimal rate of secure communication in the conditional one-time pad if an eavesdropper
has the $E$ system of the worst possible extension $\rho_{ABE}$ of the state
$\rho_{AB}$, given that squashed entanglement is defined as $\tfrac{1}{2} \inf_{\rho_{ABE}%
}\{I(A;B|E)_{\rho}:\operatorname{Tr}_{E}\{\rho_{ABE}\}=\rho_{AB}\}$
\cite{Christandl2003}. This is analogous to the interpretations from
\cite{O08}\ and\ the follow-up one in \cite{BBMW16}.

\textit{Conclusion}---In this paper, we proved that the conditional mutual
information $I(A;B|E)_{\rho}$ of a tripartite state $\rho_{ABE}$ is equal to
the optimal rate of secure communication for a task that we call the
conditional one-time pad. This represents a fundamentally different
operational interpretation of conditional mutual information that is
conceptually simple at the same time.
Furthermore, due to the fact that the optimal rate is given by conditional mutual information, the conditional one-time pad is an example of a communication task in which non-Markov quantum states are used as a resource \cite{HJPW04,EW18}. 
In the continuing quest to understand a
refined generalization of conditional mutual information, as has been
attempted previously in \cite{BSW14,DHO14,BCT16,ADJ17}, the protocol of conditional one-time pad might end up being
helpful in this effort.

\begin{acknowledgments}
We thank David Ding,  Rahul Jain, and Andreas Winter for
discussions related to the topic of this paper. MMW\ acknowledges support from the Office
of Naval Research and the National Science Foundation. KS acknowledges
support from the Department of Physics and Astronomy at LSU and the National Science
Foundation under Grant No.~1714215.
\end{acknowledgments}

\bibliographystyle{unsrt}
\bibliography{Ref}

\onecolumngrid
\appendix

\vspace{0.5in}

\begin{center}
	{\Large \bf Supplementary material}
\end{center}

\section{Preliminaries}
\label{sec:review}

We begin by reviewing some definitions and prior results relevant for the rest of the supplementary material. We point readers to \cite{W15book,H12,H06book} for background. 

\subsection{Quantum states, quantum fidelity, and trace distance}

 Throughout this work, we restrict ourselves to finite-dimensional Hilbert spaces. Let $\H$ denote a Hilbert space. Let $\D(\H)$ denote the set of density operators (positive semi-definite with unit trace) acting on $\H$. The Hilbert space for a composite system $AB$ is denoted as $\H_{AB}$ where $\H_{AB} = \H_A \otimes \H_B$. The density operator corresponding to a state of a composite system $AB$ is denoted as $\rho_{AB}$, and the reduced state $\rho_A = \tr_B\{\rho_{AB}\} $, where $\rho_A\in \D(\H_A)$. A purification of a density operator  $\rho_A$ is a pure state $\psi_{RA}$ such that $\tr_R\{\psi_{RA}\} = \rho_A$, where $R$ is called the purifying system. All purifications of a density operator are related by an isometry acting on the purifying system. The maximally mixed state acting on the Hilbert space $\H_A$ is denoted by $\pi_A \equiv I_A/\operatorname{dim}(\H_A)$.
 
 The fidelity of two quantum states $\rho,\sigma\in\mathcal{D}(\mathcal{H})$ is
 defined as \cite{U76} $F(\rho,\sigma)\equiv\left\Vert \sqrt{\rho}\sqrt{\sigma}\right\Vert _{1}^{2}$, where $\left\Vert \cdot \right \Vert_1$ denotes the trace norm. The trace distance between two density operators $\rho, \sigma \in \D(\H)$ is equal to $\left\Vert \rho - \sigma \right\Vert_1$. The operational interpretation of trace distance is that it is linearly related to the maximum success probability in distinguishing two quantum states.
 
\subsection{Typical set, strongly typicality,  typical subspace, typical projector, properties of typical subspace} \label{sec:typicality}

Suppose that a random variable $X$ takes values in an alphabet $\mathcal{X}$ with cardinality $\vert \X \vert$. Consider an i.i.d. information source that samples independently from the distribution $p_X(x)$, corresponding to random variable $X$, and emits $n$ realizations $x_1,\dots, x_n$.

Let $N(x \vert x^n)$ be the number of occurrences of the symbol $x \in \X$ in the sequence $x^n$.
\begin{definition}
	[Type]The \textit{type} or empirical distribution $t_{x^{n}}$ of a sequence
	$x^{n}$ is a probability mass function whose elements are $t_{x^{n}}( x) $
	where
	\begin{equation}
	t_{x^{n}}( x) \equiv\frac{1}{n}N(x|x^{n}).
	\end{equation}
	
\end{definition}

\begin{definition}
	[Type Class]\label{def-ct:type-class}Let $T_{t}^{X^{n}}$ denote the
	\textit{type class} of a particular
	type $t$. The type class $T_{t}^{X^{n}}$ is the set of all sequences with
	length$~n$ and type $t$:%
	\begin{equation}
	T_{t}^{X^{n}}\equiv\{x^{n}\in\mathcal{X}^{n}:t^{x^{n}}=t\}.
	\end{equation}
	
\end{definition} 

\begin{definition}
	[Strongly Typical Set]\label{def-ct:strong-typ}The $\delta$-strongly typical
	set $T_{\delta}^{X^{n}}$\ is the set of all sequences
	with an empirical distribution $\frac{1}{n}N(x|x^{n})$\ that has maximum
	deviation $\delta$ from the true distribution $p_{X}( x) $. Furthermore, the
	empirical distribution $\frac{1}{n}N(x|x^{n})$\ of any sequence in $T_{\delta
	}^{X^{n}}$\ vanishes for any letter $x$ for which $p_{X}( x) =0$:%
	\begin{equation}
	T_{\delta}^{X^{n}}\equiv\left\{  x^{n}:\forall x\in\mathcal{X},\ \left\vert
	\frac{1}{n}N(x|x^{n})-p_{X}( x) \right\vert \leq\delta\text{ if }p_{X}( x)
	>0,\text{ else }\frac{1}{n}N(x|x^{n})=0\right\}  .
	\end{equation}
	\end{definition}

We now discuss the notion of a quantum information source and recall definitions of a  typical subspace and typical projectors. Analogous to the notion of a classical information source, a quantum information source randomly emits pure qudit states in a finite-dimensional Hilbert space $\H_A$. Consider the following spectral decomposition of a density operator $\rho$:
\begin{equation}\label{eq:spec-de-rho}
\rho_A = \sum_{x\in \X} p_X(x) \ket{x} \bra{x}_A~.
\end{equation}

Now suppose that the quantum information source emits a large number $n$ of random quantum states. The density operator corresponding to the emitted state is given by
\begin{equation}
\rho_{A^n} \equiv \rho_{A_1} \otimes \dots \otimes \rho_{A^n} = \rho_A^{\otimes n}~.
\end{equation}
A spectral decomposition of the aforementioned state is as follows:
\begin{equation}
\rho_{A^n} = \sum_{x^n \in \X^n} p_{X^n}(x^n)  \ket{x^n} \bra{x^n}_{A^n}, 
\end{equation}
 where $p_{X^n}(x^n) = \prod^n_{i=1}p_X(x_i)$,  and  $\ket{x^n}_{A^n} \equiv \ket{x_1}_{A_1}\cdots \ket{x_n}_{A_n}$. 
 \begin{definition}[Typical Subspace]
 The $\delta$-typical subspace $T^{\rho, \delta}_{A^n}$ associated with many copies of a density operator, as defined in \eqref{eq:spec-de-rho},  is a subspace of the Hilbert space $\H_{A^n} = \H_{A_1}\otimes \cdots \otimes  \H_{A_n}$. It is spanned by states $\ket{x^n}_{A^n}$ whose corresponding classical sequences $x^n$ are $\delta$-typical, i.e.,
 \begin{equation}
 T^{\rho, \delta}_{A^n} \equiv \operatorname{span} \{\ket{x^n}_{A^n} : x^n \in T^{X^n}_{\delta}  \}~.
 \end{equation}
 \end{definition}
\begin{definition}[Typical Projector]
	The typical projector $\Pi^{\rho, \delta}_{A^n}$ is a projector onto the typical subspace associated with a density operator, as defined in \eqref{eq:spec-de-rho}:
	\begin{equation}
	\Pi^{\rho, \delta}_{A^n} \equiv \sum_{x^n \in T^{X^n}_{\delta}} \ket{x^n} \bra{x^n}_{A^n}.
	\end{equation}
\end{definition}


\begin{definition}
	[Type Class Subspace]The
	type class subspace associated to type $t$ is the subspace spanned by all states with the same type:
	\begin{equation}
	T_{A^{n}}^{t}\equiv\operatorname{span}\left\{  |x^{n}\rangle_{A^{n}}:x^{n}\in
	T_{t}^{X^{n}}\right\}  .
	\end{equation}
\end{definition}

\begin{definition}
	[Type Class Projector]Let $\Pi_{A^{n}}^{t}$ denote the type class subspace
	projector associated to type $t$:
	\begin{equation}
	\Pi_{A^{n}}^{t}\equiv\sum_{x^{n}\in T_{t}^{X^{n}}}|x^{n}\rangle\langle
	x^{n}|_{A^{n}}.
	\end{equation}
	
\end{definition}

Using the aforementioned definitions, we now state three useful properties of the strongly typical subspace $T^{\rho, \delta}_{A^n}$. We point readers to \cite{W15book} for a review of the proofs of these properties. 
\begin{property}[Unit Probability]
The probability that the quantum state $\rho_{A^n}$ is in the strongly typical subspace $T^{\rho, \delta}_{A^n}$ approaches one as $n$ becomes large, i.e., 
\begin{equation}
\tr\{\Pi^{\rho, \delta}_{A^n} \rho_{A^n}   \} \geq 1 - \varepsilon, 
\end{equation}
for all $\varepsilon \in (0, 1)$, $\delta > 0$, and sufficiently large $n$. 
\end{property}

\begin{property}[Exponentially Smaller Dimension]\label{prop:esd}
The dimension of the strongly typical subspace ($T^{\rho, \delta}_{A^n}$) is exponentially smaller than the dimension $\vert A\vert^n$ of the entire space of quantum states when the output of the quantum information source is not maximally mixed. The mathematical form of this property is as follows: 
\begin{equation}
\tr\{ \Pi^{\rho, \delta}_{A^n} \} \leq 2^{n (H(A) + c \delta)},
\end{equation}
where $c$ is some positive constant. 
\end{property}
\begin{property}[Equipartition]
The action of the strongly typical projector $\Pi^{\rho, \delta}_{A^n}$ on the density operator $\rho_{A^n}$ is to select out all the basis states of $\rho_{A^n}$ that are in the typical subspace and form a sliced operator. The following operator inequality holds for the sliced operator $\Pi^{\rho, \delta}_{A^n}  \rho_{A^n} \Pi^{\rho, \delta}_{A^n}$:
\begin{equation}
2^{-n (H(A) + c\delta )} \Pi^{\rho, \delta}_{A^n} \leq \Pi^{\rho, \delta}_{A^n}  \rho_{A^n} \Pi^{\rho, \delta}_{A^n} \leq 2^{-n (H(A) -c\delta )} \Pi^{\rho, \delta}_{A^n}~.
\end{equation}
\end{property}

\subsection{Packing lemma}

 \begin{definition}[Ensemble]\label{def:ensemble}
Suppose that a random variable $X$ with probability density function $p_X(x)$ takes values in an alphabet $\mathcal{X}$ with cardinality $\vert \X \vert$. Consider an ensemble $\{p_X(x), \sigma_x \}_{x\in \X}$ of quantum states where each realization x can be encoded into a quantum state $\sigma_x \in \D(\H)$. The expected density operator of the ensemble is
\begin{equation}\label{eq:ensemble}
\sigma \equiv \sum_{x \in \X} p_X(x) \sigma_x~.
\end{equation}
 \end{definition} 

For a reliable transmission of classical information, Alice can select a subset $\C$ of $\X$ for encoding, and Bob's task is to distinguish this subset of states. The subset $\C$ constitutes the code. We now recall the statement of the packing lemma. 

\begin{lemma}[Packing Lemma]\label{lem:packinglemma}
Suppose that Alice has an ensemble $\{p_X(x), \sigma_x \}_{x\in \X}$, as in Definition~\ref{def:ensemble}. Suppose that codeword subspace projectors $\{\Pi_x \}_{x\in \X}$ and a code subspace projector $\Pi$ exist, and they project onto subspaces of the Hilbert space $\H$, and these projectors and ensemble satisfy the following conditions:
\begin{align}
\operatorname{Tr}\{  \Pi\sigma_{x}\}   &  \geq1-\varepsilon
,\label{eq:pack-1}\\
\operatorname{Tr}\{  \Pi_{x}\sigma_{x}\}   &  \geq1-\varepsilon
,\label{eq:pack-2}\\
\operatorname{Tr}\{  \Pi_{x}\}   &  \leq d,\label{eq:pack-3}\\
\Pi\sigma\Pi &  \leq\frac{1}{D}\Pi, \label{eq:pack-4}%
\end{align}
where $\varepsilon\in(0,1)$, $D>0$, and $d\in(0,D)$. Suppose that
$\mathcal{M}$ is a message set of size $\left\vert \mathcal{M}\right\vert $ with
elements $m$. Consider a set $\mathcal{C}=\left\{  C_{m}\right\}
_{m\in\mathcal{M}}$ of random variables $C_{m}$ generated independently at random
according to $p_{X}(x)$, so that each random variable $C_{m}$ takes a value in
$\mathcal{X}$ and corresponds to the message $m$. However, its distribution is
independent of the particular message $m$, and therefore, the set $\mathcal{C}$\ constitutes
a random code. Then there exists a corresponding POVM\ $\{\Lambda_{m}%
\}_{m\in\mathcal{M}}$ that reliably distinguishes the states $\{\sigma_{C_{m}%
}\}_{m\in\mathcal{M}}$, in the sense that the expectation of the average
probability of detecting the correct state is high:%
\begin{equation}
\mathbb{E}_{\mathcal{C}}\left\{  \frac{1}{\left\vert \mathcal{M}\right\vert
}\sum_{m\in\mathcal{M}}\operatorname{Tr}\left\{  \Lambda_{m}\sigma_{C_{m}%
}\right\}  \right\}  \geq1-2\left(  \varepsilon+2\sqrt{\varepsilon}\right)
-4\left\vert \mathcal{M}\right\vert \frac{d}{D},
\end{equation}
when $D/d$ is large, $\left\vert \mathcal{M}\right\vert \ll D/d$, and
$\varepsilon$ is small.
\end{lemma}

We note that Bob can construct POVM\ $\{\Lambda_{m}%
\}_{m\in\mathcal{M}}$ by using the codeword subspace projectors $\{\Pi_x \}_{x\in \X}$ and the code subspace projector $\Pi$. In particular, Bob can employ a square-root measurement or a sequential decoding strategy. We point readers to \cite[Chapter 16]{W15book} for a review of an explicit construction of POVM and a complete proof of the packing lemma.  	

\subsection{Covering lemma}

The goal of the covering lemma is to cover Eve's space in such a way that Eve cannot distinguish different classical messages that Alice is sending to Bob. We start by defining two relevant ensembles for the covering lemma. We follow the convention from \cite{W15book} and refer to the two different ensembles as the ``true ensemble" and the ``fake ensemble." 

\begin{definition}[True Ensemble]
	For our discussion, the true ensemble is defined in the same way as in Definition~\ref{def:ensemble}.
\end{definition}

\begin{definition}[Fake Ensemble]
Let $\G$ be a set such that $\G \subseteq\X$. The fake ensemble is defined as follows:%
		\begin{equation}
		\left\{  1/\left\vert \G\right\vert ,\sigma_{g}\right\}
		_{g\in\G}.
		\end{equation}
		Let $\overline{\sigma}$ denote the ``fake expected density
		operator'' of the fake ensemble:%
		\begin{equation}\label{eq:fake-ensemble}
		\overline{\sigma}(\G)\equiv\frac{1}{\left\vert \G\right\vert
		}\sum_{g\in\G}\sigma_g.
		\end{equation}
\end{definition}

The goal for Alice is to generate a fake ensemble such that the trace distance between $\bar{\sigma}(\G)$ in \eqref{eq:fake-ensemble} and $\sigma$ in \eqref{eq:ensemble} is small. 
Moreover, in order to achieve a higher private communication rate, it is required to make the size of fake ensembles as small as possible while still having privacy from Eve. We call the trace distance between $\bar{\sigma}(\G)$ and $\sigma$ the obfuscation error $o_{e}(\mathcal{\G})$, i.e., 
\begin{equation}
o_{e}(\mathcal{\G}) = \left \Vert \bar{\sigma}(\G) - \sigma \right \Vert_{1}~.
\end{equation}

We now recall the statement of the covering lemma. 
\begin{lemma}
	[Covering Lemma]
	\label{lemma-cov:covering}Let $\left\{  p_{X}(x),\sigma
	_{x}\right\}  _{x\in\mathcal{X}}$ be an ensemble as in
	Definition~\ref{def:ensemble}. Suppose a total subspace projector $\Pi$
	and codeword subspace projectors $\left\{  \Pi_{x}\right\}  _{x\in\mathcal{X}%
	}$ are given, they project onto subspaces of $\mathcal{H}$, and these projectors
	and each state $\sigma_{x}$ satisfy the following conditions:%
	\begin{align}
	\operatorname{Tr}\{  \Pi \sigma_{x}\}   &  \geq1-\varepsilon
	,\label{eq-cov:cov-1}\\
	\operatorname{Tr}\{  \Pi_{x} \sigma_{x} \}   &  \geq1-\varepsilon
	,\label{eq-cov:cov-2}\\
	\operatorname{Tr}\{  \Pi\}   &  \leq D,\label{eq-cov:cov-3}\\
	\Pi_{x}\sigma_{x}\Pi_{x}  &  \leq\frac{1}{d}\Pi_{x}, \label{eq-cov:cov-4}%
	\end{align}
	where $\varepsilon\in(0,1)$, $D>0$, and $d\in(0,D)$. Suppose that
	$\G$ is a set of size $\left\vert \G \right\vert $ with
	elements $g$. Let a random covering code $\mathcal{C}\equiv\left\{
	C_{g}\right\}  _{g\in\mathcal{G}}$ consist of random codewords $C_{g}$
	where\ the codewords $C_{g}$ are chosen independently according to the distribution
	$p_{X}(x)$ and give rise to a fake ensemble $\left\{  1/\left\vert
	\mathcal{G}\right\vert ,\sigma_{C_{g}}\right\}  _{g\in\mathcal{G}}$. Then the following bound exists on the probability of having a small obfuscation error $o_e(\C)$ of the random covering code:
	\begin{equation}
	\Pr_{\mathcal{C}}\left\{  o_{e}(\mathcal{C})\leq\varepsilon+4\sqrt
	{\varepsilon}+24\sqrt[4]{\varepsilon}\right\}  \geq1-2D\exp\left(
	-\frac{\varepsilon^{3}}{4}\frac{\left\vert \mathcal{G}\right\vert d}%
	{D}\right)  ,
	\end{equation}
	when $\varepsilon$ is small and $\left\vert \mathcal{G}\right\vert
	\gg\varepsilon^{3}d/D$. Thus, it is highly likely that a given fake ensemble
	$\left\{  1/\left\vert
	\mathcal{G}\right\vert ,\sigma_{C_{g}}\right\}  _{g\in\mathcal{G}}$\ has its expected density operator indistinguishable from
	the expected density operator of the original ensemble $\left\{
	p_{X}(x),\sigma_{x}\right\}  _{x\in\mathcal{X}}$.
\end{lemma}

We point readers to \cite[Chapter 17]{W15book} for a proof of the covering lemma.  
 
\subsection{Properties of encoding unitaries} 

\label{sec:encoding-unitaries}

Consider the reduced state $\rho_{A}$ and a
spectral decomposition for it as $\rho_{A}=\sum_{x}p_{X}(x)|x\rangle\langle
x|_{A}$, where $p_{X}$ is a probability distribution and $\{|x\rangle
_{A}\}_{x}$ is an orthonormal basis. Consider the following purification of $\rho_A$:
\begin{equation}
\vert \psi\rangle_{AR} = \sum_x \sqrt{p_X(x)} \vert x \rangle_A \vert x\rangle_R, 
\end{equation} 
where $p_X(x) >0$ for all $x$, $\sum_x p_X(x) = 1$, and $\{\vert x\rangle_A \}$ and $\{\vert x\rangle_R \}$ are orthonormal bases for systems $A$ and $R$, respectively. We now start with $n$ copies of the above state and write in terms of its type decomposition, as given in Definition~\ref{def-ct:type-class}:
\begin{align}
\ket{\psi^n}_{A^nR^n} &= \sum_{x^n} \sqrt{p_{X^n}(x^n)} \ket{x^n}_{A^n}\ket{x^n}_{R^n}\\
&  =\sum_{t}\sqrt{p_{X^{n}}(x_{t}^{n})d_{t}}\frac{1}{\sqrt{d_{t}}}\sum
_{x^{n}\in T_{t}}|x^{n}\rangle_{A^{n}}|x^{n}\rangle_{R^{n}}\\
&  =\sum_{t}\sqrt{p(t)}|\Phi_{t}\rangle_{A^{n}R^{n}},
\label{eq-eac:type-decomp-4}
\end{align}
where $d_{t}$ is the size of the type class $T_t$ and 
\begin{align}
p(t)  &  \equiv p_{X^{n}}(x_{t}^{n})d_{t},\label{eq-eac:pt-dist}\\
|\Phi_{t}\rangle_{A^{n}R^{n}}  &  \equiv\frac{1}{\sqrt{d_{t}}}\sum_{x^{n}\in
T_{t}}|x^{n}\rangle_{A^{n}}|x^{n}\rangle_{R^{n}}.
\end{align}

We now consider a Heisenberg-Weyl set of $d^2_t$ operators that act on all the $A^n$ systems of $\ket{\Phi_t}_{A^nR^n}$. We denote one of these operators by $V(x_{t}
,z_{t})\equiv X(x_{t})Z(z_{t})$ where $x_{t},z_{t}\in\left\{  0,\ldots,d_{t}-1\right\}$. Along with these operators, Alice applies a phase $(-1)^{b_t}$ in each subspace. Therefore, the resulting unitary operator can be expressed as a direct sum of all these unitary operators:
 \begin{equation}\label{eq:unitary-operator}
	U_{A^{n}}(s)=\bigoplus\limits_{t}(-1)^{b_{t}}V_{A^{n}}(x_{t},z_{t}), 
\end{equation}
where $s$ is a vector containing all the indices needed to specify the unitary $U(s)$:
\begin{equation}\label{eq:s-variable}
	s \equiv [(x_t, z_t, b_t)]_t.
\end{equation}

We now recall that a transpose trick holds for such unitary operators. Consider the following chain of inequalities:
\begin{align}
(U_{A^{n}}(s)\otimes I_{R^n}) \ket{\psi^n}_{A^n R^n} &=\sum_{t}\sqrt{p(t)} (-1)^{b_{t}}V_{A^{n}}(x_{t},z_{t})\ket{\Phi_{t}}_{A^{n}R^{n}}\\
& =\sum_{t}\sqrt{p(t)} (-1)^{b_{t}}V_{R^{n}}^{T}(x_{t},z_{t})\ket{\Phi_{t}}_{A^{n}R^{n}}\\
& =\left(  I_{A^{n}}\otimes U_{R^{n}}^{T}(s)\right)  \ket{\psi^n}_{A^{n}R^{n}},
\end{align}
where we have used the direct-sum property of the unitary operator \eqref{eq:unitary-operator} and a transpose trick associated with the maximally entangled state $\ket{\Phi_t}_{A^n R^n}$.

Let $\ket{\psi}_{ABEF}$ denote a purification of $\rho_{ABE}$. Since all purifications are related by an isometry acting on the purifying system, there exists an isometry
$U_{R\rightarrow BEF}$ such that $U_{R\rightarrow BEF}|\phi\rangle_{AR}%
=|\psi\rangle_{ABEF}$. Applying the isometry $U_{R\rightarrow BEF}$ followed
by a partial trace over $F$ can be thought of as a channel $\mathcal{N}%
_{R\rightarrow BE}$ that realizes the state $\rho_{ABE}$ as $\mathcal{N}%
_{R\rightarrow BE}(\phi_{AR})=\rho_{ABE}$. Moreover, the state $\rho_{ABE}^{\otimes n}$
can be thought of as the following state $\mathcal{N}_{R\rightarrow
	BE}^{\otimes n}(\phi_{AR}^{\otimes n})$.

\section{Achievability of CQMI for conditional one-time pad}\label{sec:achievability-of-CQMI}

In this section, we provide a proof that the conditional quantum mutual information $I(A;B \vert E)_{\rho}$ is a lower bound on the conditional one-time pad capacity of $\rho_{ABE}$.

In the conditional one-time pad task, the goal for Alice is to encode information in her share of the state $\rho_{ABE}$ in such a way that Bob can reliably decode the information, while maintaining privacy from Eve. We now construct a coding scheme based on random coding for reliable communication between Alice and Bob. Let $M, K \in \mathbb{N}$. Alice has message variable $m\in \{1, \dots, M \}$ and a local key variable $k \in \{1, \dots, K \}$. If Alice wishes to send message $m$, then she picks $k$ uniformly at random from $k \in \{1, \dots, K \}$. For each pair $(m, k)$, the random code is selected in such a way that the vector $s$, of the form described in \eqref{eq:s-variable}, is chosen uniformly at random and then the encoding unitary $U_{A^n}(s(m,k))$ acting on the state $\rho^{\otimes n}_{ABE}$ is associated to $m$ and $k$. Therefore, the set $\C = \{s(m, k)\}_{m, k }$ constitutes a random code, given that the $s$ vectors are picked uniformly at random; i.e., the way that they are selected is  independent of $m$ and $k$, but after they are chosen, the association to $m$ and $k$ is made. Let $\S$ denote the set of all possible vectors $s$. 
To be clear, the ensemble from which Alice and Bob are selecting their code can be expressed as
\begin{equation}\label{eq:packing-lemma-ensemble}
\left\{\frac{1}{\vert \S \vert}, U_{A^n}(s)\rho_{A^nB^nE^n}U_{A^n}^{\dag}(s) \right\}_{s \in \S}.
\end{equation}

As described in Lemma \ref{lem:packinglemma}, if the four inequalities corresponding to the codeword subspace projectors, the code subspace projector, and aforementioned ensemble are satisfied, then there exists a decoding POVM that can reliably decode Alice's transmitted message. Consider the following respective codeword subspace projectors and a code subspace projector:
\begin{align}
&  U_{A^n}(s)\Pi_{A^nB^nE^n}^{\rho,\delta}U_{A^n}^{\dag}(s),\\
&  \Pi_{A^n}^{\rho,\delta}\otimes\Pi_{B^nE^n}^{\rho,\delta},
\end{align}
where $\Pi_{A^nB^nE^n}^{\rho,\delta}$, $\Pi_{A^n}^{\rho,\delta}$, and $\Pi_{B^nE^n}^{\rho,\delta}$ are the typical projectors for $n$ copies of the states $\rho_{ABE}$, $\rho_A$, and $\rho_{BE}$, respectively. 

Let $\bar{\rho}_{A^nB^nE^n}$ denote the expected density operator of the ensemble in $\eqref{eq:packing-lemma-ensemble}$. We now state the four conditions corresponding to the packing lemma for our code:
\begin{align}
\tr \left\{\left(  \Pi_{A^n}^{\rho,\delta}\otimes \Pi_{B^nE^n}^{\rho,\delta}\right)  \left(U_{A^n}(s)\rho_{A^nB^nE^n}U_{A^n}^{\dag}(s)\right)\right\}   &  \geq1-\varepsilon
,\label{eq:plemma1}\\
\tr\left\{  \left(  U_{A^n}(s)\Pi_{A^nB^nE^n}^{\rho,\delta}U_{A^n}^{\dag}(s)\right) \left(  U_{A^n}(s)\rho_{A^nB^nE^n}U_{A^n}^{\dag}(s)\right)  \right\}   &\geq1-\varepsilon,\label{eq:plemma2}\\
\tr\left\{ U_{A^n}(s)\Pi_{A^nB^nE^n}^{\rho
,\delta}U_{A^n}^{\dag}(s)\right\}   &  \leq2^{n\left[  H(ABE)_{\rho
}+c\delta\right]  } \label{eq:plemma3}\\
\left(  \Pi_{A^n}^{\rho,\delta}\otimes \Pi_{B^nE^n}^{\rho,\delta}\right)  \overline{\rho}_{A^n B^n E^n}\left(  \Pi_{A^n}^{\rho,\delta}\otimes \Pi_{B^n E^n}^{\rho,\delta}\right) 
&\leq2^{-n\left[  H(A)_{\rho}+H(BE)_{\rho}-\nu(n,\delta)-c\delta
\right]  }\left(  \Pi_{A^{ n}}^{\rho,\delta}\otimes\Pi_{B^{n} E^n}^{\rho,\delta}\right)\label{eq:plemma4},
\end{align}
where $c$ is some constant, and $\nu(n,\delta)$ is given by
\begin{align}
\nu(n, \delta) =  \frac{\delta}{2} \operatorname{dim}(\H_{B}) \log_2\operatorname{dim}(\H_{B}) + h_2\!\left(\operatorname{dim}(\H_{B}) \frac{\delta}{2}\right) + \operatorname{dim}(\H_{B})\frac{1}{n}\log(n+1) ,
\end{align}
which approaches zero as $n\to \infty$ and $\delta\to 0$. In order to establish these inequalities, we use the properties of typical projectors described in Section \ref{sec:typicality} and encoding unitaries described in Section \ref{sec:encoding-unitaries}. We point readers to \cite[Chapter 23]{W15book} for a review of related proofs to establish these inequalities.

We now invoke Lemma \ref{lem:packinglemma} to demonstrate the existence of a reliable code. Since the four conditions \eqref{eq:plemma1}--\eqref{eq:plemma4} hold, there exists a POVM $\{\Lambda^{m,k}_{A^nB^nE^n}\}_{m , k }$ that can detect the transmitted states with an arbitrarily low expectation of the average probability of error, as described in Lemma \ref{lem:packinglemma}. In particular, we get the following upper bound on the expectation of average probability of error:
\begin{align}
\mathbb{E}_{\mathcal{C}}&\Bigg\{\frac{1}{MK}\sum_{m,k}\tr\{(I - \Lambda_{A^{n}B^{n}E^{n}}^{m,k}) 
U_{A^{n}}(S(m,k))\rho_{ABE}^{\otimes n} U_{A^{n}}^{\dag}(S(m,k))\}\Bigg\} \notag \\
&\leq 2\left(  \varepsilon+2\sqrt{\varepsilon}\right)
+4~M K ~2^{-n\left[  H(A)_{\rho}+H(BE)_{\rho}-\nu(n,\delta)-c\delta
\right]  }2^{n\left[  H(ABE)_{\rho
}+c\delta\right]  } \\
&\leq 2\left(  \varepsilon+2\sqrt{\varepsilon}\right)
+4~M K ~2^{-n\left[  I(A;BE)_{\rho}-\nu(n,\delta)-2c\delta
\right]}\\
& \leq 2\left(  \varepsilon+2\sqrt{\varepsilon}\right) + 4\cdot2^{-n c \delta} \label{eq:expectation-avg-error},
\end{align}
where we considered the size of the message set and the local key set combined to be 
\begin{align}
M K = 2^{n\left[ I(A;BE) - \nu(n, \delta) - 3 c\delta \right]}~.
\end{align}

Therefore, the number of bits per channel use encoded by $M$ and $K$ is
\begin{align}
\frac{1}{n} \log_2 M K = I(A;BE)_{\rho} -  \nu(n, \delta) - 3 c \delta~.
\end{align}

Let $\varepsilon' \in (0,1)$ and $\delta' \in (0,1)$. If we pick $n$ large enough and $\delta$ small enough, we can have both $\nu(n, \delta) + 3 c \delta \leq \delta'$ and  $2\left(  \varepsilon+2\sqrt{\varepsilon}\right) + 4\cdot2^{-n c \delta} \leq \varepsilon'$. Therefore, if $\log_2 M K \approx n I(A;BE)_{\rho}$, Alice can reliably communicate classical messages to Bob.

We now provide a proof for maintaining privacy from an eavesdropper in the conditional one-time pad task. Consider the following respective codeword subspace projectors and a code subspace projector.
\begin{align}
&  U_{A^n}(s)\Pi_{A^nE^n}^{\rho,\delta}U_{A^n}^{\dagger}(s),\\
&  \Pi_{A^n}^{\rho,\delta}\otimes\Pi_{E^n}^{\rho,\delta},
\end{align}
where $\Pi_{A^nE^n}^{\rho,\delta}$, $\Pi_{A^n}^{\rho,\delta}$, and $\Pi_{E^n}^{\rho,\delta}$ are the typical projectors for many copies of the states $\rho_{AE}$, $\rho_A$, and $\rho_{E}$, respectively. 

Furthermore, consider the following ensemble derived from \eqref{eq:packing-lemma-ensemble} by tracing over the $B^n$ systems:
\begin{align}
\left\{\frac{1}{\vert \S \vert}, U_{A^n}(s)\rho_{A^nE^n}U_{A^n}^{\dagger}(s) \right\}_{s \in \S}.
\end{align}
The ensemble average of this ensemble is given by
\begin{equation}\label{eq:expected-eve-state}
\overline{\tau}_{A^{n}E^{n}}\equiv\mathbb{E}_{S}\left\{  U_{A^{n}}(S)\rho
_{AE}^{\otimes n}U_{A^{n}}^{\dag}(S)\right\}  .
\end{equation}

Since for any message $m$, Alice picks $k$ uniformly at random from $k \in \{1, \dots , K \}$, then from the perspective of an Eve who does not know the choice of $k$, the state has the following form:
\begin{align}\label{eq:eve-state}
\tau_{A^{n}E^{n}}^{m}\equiv\frac{1}{K}\sum_{k=1}^{K}U_{A^{n}}(s)\rho_{AE}^{\otimes n}U_{A^{n}}^{\dag}(s).
\end{align}

As described in Lemma \ref{lemma-cov:covering}, if the four inequalities corresponding to the codeword subspace projectors, the code subspace projector, and the above mentioned ensemble are satisfied, then it is highly likely that $\tau_{A^{n}E^{n}}^{m}$ in \eqref{eq:eve-state} is indistinguishable from  $\overline{\tau}_{A^{n}E^{n}}$ in \eqref{eq:expected-eve-state}.

We now state the four conditions corresponding to the covering lemma for our code:
\begin{align}
\tr\{( \Pi_{A^n}^{\rho,\delta}\otimes\Pi_{E^n}^{\rho,\delta}) (U_{A^n}(s)\rho_{A^nE^n}U_{A^n}^{\dagger}(s)) \} &\geq 1 - \varepsilon,\\
\tr\{(U_{A^n}(s)\Pi_{A^nE^n}^{\rho,\delta}U_{A^n}^{\dagger}(s)) (U_{A^n}(s)\rho_{A^nE^n}U_{A^n}^{\dagger}(s)) \} &\geq 1-\varepsilon,\\
\tr\{  \Pi_{A^n}^{\rho,\delta}\otimes\Pi_{E^n}^{\rho,\delta}\} &\leq 2^{n (H(A)_{\rho}+ H(E)_{\rho} + 2 c \delta)},\\
\left(U_{A^n}(s)\Pi_{A^nE^n}^{\rho,\delta}U_{A^n}^{\dagger}(s) \right) U_{A^n}(s)\rho_{A^nE^n}U_{A^n}^{\dagger}(s) \left(U_{A^n}(s)\Pi_{A^nE^n}^{\rho,\delta}U_{A^n}^{\dagger}(s)\right)&\leq 2^{-n(H(AE)_\rho - c \delta) }\left(U_{A^n}(s)\Pi_{A^nE^n}^{\rho,\delta}U_{A^n}^{\dagger}(s)\right)~,
\end{align}
where $c$ is some constant. Proofs of these properties are available in \cite{W15book}.

We now invoke Lemma \ref{lemma-cov:covering} and arrive at the following inequality:
\begin{align}
\Pr_{\mathcal{C}}\{\left\Vert \tau_{A^{n}E^{n}}^{m}-\overline{\tau}_{A^{n}E^{n}}\right\Vert
_{1}\leq\varepsilon+4\sqrt{\varepsilon}+24\sqrt[4]{\varepsilon}\}
\geq1-2^{n(H(A)_{\rho}+H(E)_{\rho}+2 c\delta+1/n)}\exp\left(  -\frac{\varepsilon^{3}K2^{-n[I(A;E)_{\rho}- 3 c \delta]}}{4}\right)~.
\end{align}
Thus, if we choose the size of the key set to be  $K = 2^{n[I(A;E)_{\rho} +4 c \delta]}$, then $\exp\{-\varepsilon^3 2^{n c \delta} /(4 )  \}$ is doubly exponentially decreasing in $n$. Therefore, if $\log_2 K \approx n I(A;E)_{\rho}$, it is highly likely that the state $\tau_{A^{n}E^{n}}^{m}$ will be nearly
indistinguishable from the average state $\overline{\tau}_{A^{n}E^{n}}$.

As described earlier, in the conditional one-time pad task, the goal for Alice is to encode information in her share of the state $\rho_{ABE}$ in such a way that Bob can reliably decode the information, while maintaining privacy from Eve. So far, we have shown that Alice can reliably communicate to Bob. Moreover, we have also discussed a strategy that Alice can implement to communicate a classical message $m$ to Bob, such that the quantum state that Eve can access has essentially no dependence on the message $m$.

Next, we would like to show the existence of a code that is both reliable and secure. Using the
union bound of probability theory, it can be shown that there is a non-zero probability for there to be a
code $\{s(m,k)\}_{m,k}$ such that the average success probability of
Bob's decoder is arbitrarily high and $\left\Vert \tau_{A^{n}E^{n}}%
^{m}-\overline{\tau}_{A^{n}E^{n}}\right\Vert _{1}$ is arbitrarily small for
all $m$, with these statements holding for sufficiently large $n$. 
Furthermore, a final ``expurgation" argument can be applied to show that Bob can decode each $m$ and $k$ with arbitrarily high probability and that $\left\Vert \tau_{A^{n}E^{n}}%
^{m}-\overline{\tau}_{A^{n}E^{n}}\right\Vert _{1}$ is arbitrarily small for
all $m$. These techniques have been used to establish a formula for the capacity of quantum channel for transmitting private classical information in \cite{ieee2005dev,1050633}. We point readers to \cite[Chapter 23]{W15book} for a review of a related proof to establish the desired result. 

Therefore, the number of bits that Alice can communicate securely is
\begin{align}
\log_{2}M=\log_{2}MK-\log_{2}K\approx n[I(A;BE)_{\rho}-I(A;E)_{\rho}]=nI(A;B|E)_{\rho}~,
\end{align}
and $I(A;B|E)_{\rho}$ is an achievable rate. This
concludes the achievability proof.

\section{Optimality of CQMI for conditional one-time pad}\label{sec:optimality-of-CQMI}

In this appendix, we establish
that the conditional one-time pad capacity of $\rho_{ABE}$ cannot exceed
$I(A;B|E)_{\rho}$. To see this, consider an arbitrary $(n,M,\varepsilon
,\delta)$\ protocol of the above form, and suppose that the message $m$ is
chosen uniformly at random. Then the overall state that describes all systems
is%
\begin{equation}
\omega_{\hat{M}A^{\prime}B^{n}E^{n}}\equiv\frac{1}{M}\sum_{m=1}^{M}%
|m\rangle\langle m|_{\hat{M}}\otimes\omega_{A^{\prime}B^{n}E^{n}}^{m}, \label{eq:init-combined-state}%
\end{equation}
where $\{|m\rangle_{\hat{M}}\}_{m}$ is an orthonormal basis and the state $\omega_{A^{\prime}B^{n}E^{n}}^{m}$ is defined in the main text. We can describe
Bob's decoding measurement as a measurement channel
\begin{equation}
\mathcal{M}_{A^{\prime
}B^{n}E^{n}\rightarrow M^{\prime}}(\theta_{A^{\prime}B^{n}E^{n}})\equiv
\sum_{m}\operatorname{Tr}\{\Lambda_{A^{\prime}B^{n}E^{n}}^{m}\theta
_{A^{\prime}B^{n}E^{n}}\}|m\rangle\langle m|_{M^{\prime}},\label{eq:measurement-channel}%
\end{equation}
so that the final
output state is%
\begin{equation}
\omega_{\hat{M}M^{\prime}}=\mathcal{M}_{A^{\prime}B^{n}E^{n}\rightarrow
M^{\prime}}(\omega_{\hat{M}A^{\prime}B^{n}E^{n}}).
\end{equation}
By the condition in \eqref{eq:reliability-cond} and some further calculations,
it follows that%
\begin{equation}
\frac{1}{2}\left\Vert \omega_{\hat{M}M^{\prime}}-\overline{\Phi}_{\hat{M}M^{\prime}%
}\right\Vert _{1}\leq\varepsilon, \label{eq:c4}
\end{equation}
where $\overline{\Phi}_{\hat{M}M^{\prime}}\equiv\frac{1}{M}\sum_{m=1}%
^{M}|m\rangle\langle m|_{\hat{M}}\otimes|m\rangle\langle m|_{M^{\prime}}$ is a
maximally classically correlated state. A uniform bound for the continuity of
mutual information \cite{Winter15}\ implies that%
\begin{align}
\log_{2}M  & =I(\hat{M};M^{\prime})_{\overline{\Phi}}\\
& \leq I(\hat{M};M^{\prime})_{\omega}+\varepsilon\log_2 M+g(\varepsilon),
\end{align}
where $g(\varepsilon)\equiv(\varepsilon+1)\log_{2}(\varepsilon+1)-\varepsilon
\log_{2}\varepsilon$, with the property that $\lim_{\varepsilon\rightarrow
0}g(\varepsilon)=0$. From the Holevo
bound \cite{Holevo73} or more generally quantum data processing (see, e.g., \cite[Section~11.9.2]{W15book}), it follows that%
\begin{equation}
I(\hat{M};M^{\prime})_{\omega}\leq I(\hat{M};A^{\prime}B^{n}E^{n})_{\omega}.
\end{equation}
By the condition in \eqref{eq:security-cond}, it follows that%
\begin{equation}
\frac{1}{2}\left\Vert \omega_{\hat{M}A^{\prime}E^{n}}-\omega_{\hat{M}}%
\otimes\sigma_{A^{\prime}E^{n}}\right\Vert _{1}\leq\delta,
\end{equation}
which in turn implies from \cite{Winter15}\ that%
\begin{equation}
I(\hat{M};A^{\prime}E^{n})_{\omega}\leq\delta\log_2 M+g(\delta).
\end{equation}
Putting everything together leads to the following bound:%
\begin{align}
\log_{2}M\leq I(\hat{M};B^{n}|A^{\prime}E^{n})_{\omega}
+\left(  \varepsilon+\delta\right)  \log_2 M+g(\varepsilon)+g(\delta),\label{eq:c10}
\end{align}
where we used that $I(\hat{M};A^{\prime}B^{n}E^{n})_{\omega}-I(\hat
{M};A^{\prime}E^{n})_{\omega}=I(\hat{M};B^{n}|A^{\prime}E^{n})_{\omega}$. Now
by several applications of the chain rule for conditional mutual information,
we find that%
\begin{align}
  I(\hat{M};B^{n}|A^{\prime}E^{n})_{\omega}  
 & =\sum_{i=1}^{n}I(\hat{M}%
;B_{i}|B^{i-1}A^{\prime}E^{n})_{\omega}
\label{eq:cmi-bnd-first}
\\
& \leq\sum_{i=1}^{n}I(\hat{M}A^{\prime}B^{i-1}E^{i-1}E^{n}_{i+1};B_{i}|E_{i})_{\omega}\\
& \leq\sum_{i=1}^{n}I(A_{i};B_{i}|E_{i})_{\rho^{\otimes n}}\\
& =nI(A;B|E)_{\rho}. \label{eq:cmi-bnd-final}
\end{align}
The second inequality follows because we can consider the sequential action of
1)\ tensoring in the states $\rho_{ABE}^{\otimes i-1}$ and $\rho
_{ABE}^{\otimes n-i}$ to the $i$th copy of $\rho_{ABE}$, 2)\ tensoring in the state $\frac{1}{M}\sum_{m=1}%
^{M}|m\rangle\langle m|_{\hat{M}}$, 3)\ applying the encoding $\mathcal{E}%
_{A^{n}\rightarrow A^{\prime}}^{m}$ conditioned on the value $m$ in $\hat{M}$,
and 4)\ tracing over the systems $B_{i+1}^{n}$ all as a local
channel $\mathcal{N}_{A_{i}\rightarrow\hat{M}A^{\prime}B^{i-1}E^{i-1}E_{i+1}^n}$ acting
on the $A_{i}$ system of $\rho_{A_{i}B_{i}E_{i}}$, so that%
\begin{equation}
\omega_{\hat{M}A^{\prime}B^{i-1}E^{i-1}B_{i}E_{i}}=\mathcal{N}_{A_{i}%
\rightarrow\hat{M}A^{\prime}B^{i-1}E^{i-1}E_{i+1}^n}(\rho_{A_{i}B_{i}E_{i}}),
\end{equation}
and the conditional mutual information does not increase under the action of a
local channel on an unconditioned system \cite{Christandl2003}:
\begin{equation}
I(\hat
{M}A^{\prime}B^{i-1}E^{i-1};B_{i}|E_{i})_{\omega}\leq I(A_{i};B_{i}%
|E_{i})_{\rho^{\otimes n}}.
\end{equation}

Alternatively, the inequality resulting from \eqref{eq:cmi-bnd-first}--\eqref{eq:cmi-bnd-final} may be seen by the following steps:
\begin{align}
I(\hat{M};B^{n}|A^{\prime}E^{n})_{\omega}
& = I(\hat{M}A^{\prime};B^{n}|E^{n})_{\omega}
- I(A^{\prime};B^{n}|E^{n})_{\omega} \\
& \leq I(\hat{M}A^{\prime};B^{n}|E^{n})_{\omega}
\\
& \leq I(A^n;B^{n}|E^{n})_{\rho^{\otimes n}}\\
& = n I(A;B|E)_{\rho}.
\end{align}
The first inequality follows from non-negativity of conditional mutual information \cite{LR73,PhysRevLett.30.434}, and the second inequality follows from monotonicity of conditional mutual information \cite{Christandl2003} with respect to a local channel acting on one of the unconditioned systems [in this case, the local channel is the encoding channel that tensors in the maximally mixed state on system $\hat{M}$ and applies the channel $\sum_m \vert m \rangle \langle m \vert_{\hat{M}}(\cdot) \vert m \rangle \langle m \vert_{\hat{M}} \otimes \mathcal{E}%
_{A^{n}\rightarrow A^{\prime}}^{m}(\cdot)$].

Putting everything together, we find the
following bound for any $(n,M,\varepsilon,\delta)$ conditional one-time pad
protocol:%
\begin{equation}
\frac{1-\varepsilon-\delta}{n}\log_{2}M\leq I(A;B|E)_{\rho}+\frac
{g(\varepsilon)+g(\delta)}{n}.
\end{equation}
Taking the limit as $n\rightarrow\infty$ and then as $\varepsilon
,\delta\rightarrow0$ allows us to conclude that the conditional mutual
information $I(A;B|E)_{\rho}$ is an upper bound on the conditional one-time
pad capacity of $\rho_{ABE}$.

\section{A proof of the converse theorem for a secret-sharing task}

In this section, we provide a proof of the converse theorem for a secret-sharing task that we call \textit{information scrambling}. The goal of the information scrambling task is for Alice to communicate a message in such a way that it can be decoded only by someone who possesses all three $ABE$ systems. If someone possesses only the $AB$ systems or only the $AE$ systems, then such a person can figure out essentially nothing about the encoded message. 

By using arguments similar to \eqref{eq:c4}--\eqref{eq:c10}, we find the following two inequalities:
\begin{align}
\log_{2}M&\leq n I(A;B \vert E)_{\rho}
+\left(  \varepsilon+\delta\right)  \log_2 M+g(\varepsilon)+g(\delta)~. \label{eq:D1} \\ 
\log_{2}M &\leq n I(A;E \vert B)_{\rho}
+\left(  \varepsilon+\delta\right)  \log_2 M+g(\varepsilon)+g(\delta)~. \label{eq:D2}
\end{align}
Therefore, by combining \eqref{eq:D1} and \eqref{eq:D2}, we arrive at the following bound for any $(n, M, \varepsilon, \delta)$ secret-sharing protocol:%
\begin{equation}
\frac{1-\varepsilon-\delta}{n}\log_{2}M\leq I(A;BE)_{\rho} - \max\{I(A;B)_{\rho}, I(A;E)_{\rho} \}+\frac
{g(\varepsilon)+g(\delta)}{n}.
\end{equation}
Taking the limit as $n\rightarrow\infty$ and then as $\varepsilon
,\delta\rightarrow0$ allows us to conclude that $I(A;BE)_{\rho} - \max\{I(A;B)_{\rho}, I(A;E)_{\rho}\}$ is an upper bound on the information scrambling capacity of $\rho_{ABE}$.

The proof for the achievability part follows similarly to the achievability part for the conditional one-time pad task, except that we have extra security conditions that should hold. To handle this, we just invoke the covering lemma again to be sure that the key variable is large enough to protect the message variable against local parties who do not have access to all systems of the full state.

\section{A proof of the converse theorem for the communication protocol in \cite{BO12}}

We begin by recalling the communication protocol described in \cite{BO12}. Suppose that Alice, Bob, and Eve share $n$~copies of the quantum state $\rho_{ABE}$, so that their state is $\rho^{\otimes n}_{ABE}$. In this communication protocol, Alice, Bob, and Eve have access to the $A$, $B$, and $E$ systems, respectively. Alice and Bob are connected by an ideal quantum channel, which Eve has access to as well. The goal of this protocol is for Alice to encode a message $m$ into her $A$ systems, in such a way that if she sends her $A$ systems over the ideal quantum channel, then Bob can decode the message $m$ reliably by performing a measurement on all of the $AB$ systems, while Eve, possessing the $AE$ systems, has essentially no chance of determining the message $m$ if she tried to figure it out. 

At the beginning of the protocol, Alice picks $m\in\{1,\ldots,M\}$ and applies an encoding channel $\mathcal{E}%
_{A^{n}\rightarrow A^{\prime}}^{m}$ to the $A^{n}$ systems of $\rho
_{ABE}^{\otimes n}$, leading to the state $\omega_{A^{\prime}B^{n}E^{n}}%
^{m}\equiv\mathcal{E}_{A^{n}\rightarrow A^{\prime}}^{m}(\rho_{ABE}^{\otimes
n})$. She transmits the system $A^{\prime}$ of $\omega_{A^{\prime}B^{n}E^{n}%
}^{m}$ over the ideal quantum channel. Bob applies a decoding positive
operator-valued measure $\{\Lambda_{A^{\prime}B^{n}}^{m}\}_{m}$ to the
systems $A^{\prime}B^{n}$ of $\omega_{A^{\prime}B^{n}E^{n}}^{m}$ in order
to figure out which message was transmitted. The protocol is $\varepsilon
$-reliable if Bob can determine the message $m$ with probability not smaller
than $1-\varepsilon$:%
\begin{equation}
\forall m:\operatorname{Tr}\{\Lambda_{A^{\prime}B^{n}}^{m}\omega
_{A^{\prime}B^{n}}^{m}\}\geq1-\varepsilon.%
\end{equation}
The protocol is $\delta$-secure if the reduced state $\omega_{A^{\prime}E^{n}%
}^{m}$\ on systems $A^{\prime}E^{n}$ is nearly indistinguishable from a
constant state $\sigma_{A^{\prime}E^{n}}$ independent of the message $m$:%
\begin{equation}
\forall m:\frac{1}{2}\left\Vert \omega_{A^{\prime}E^{n}}^{m}-\sigma
_{A^{\prime}E^{n}}\right\Vert _{1}\leq\delta~. %
\end{equation}
Achievable rates and capacity are defined similarly to the previous cases.

We now provide a proof to establish an upper bound on the rate of communication for an arbitrary protocol of the above form, which is different from the proof given in \cite{BO12}. Let the message $m$ be chosen uniformly at random, and the overall state that describes all systems is defined in the same way as in \eqref{eq:init-combined-state}. We can describe Bob's decoding measurement as a measurement channel similar to \eqref{eq:measurement-channel}:
\begin{equation}
\mathcal{M}_{A^{\prime
}B^{n}\rightarrow M^{\prime}}(\theta_{A^{\prime}B^{n}})\equiv
\sum_{m}\operatorname{Tr}\{\Lambda_{A^{\prime}B^{n}}^{m}\theta
_{A^{\prime}B^{n}}\}|m\rangle\langle m|_{M^{\prime}},
\end{equation}
so that the final
output state is%
\begin{equation}
\omega_{\hat{M}M^{\prime}}=\mathcal{M}_{A^{\prime}B^{n}\rightarrow
M^{\prime}}(\omega_{\hat{M}A^{\prime}B^{n}}),
\end{equation}
where
\begin{equation}
\omega_{\hat{M}A^{\prime}B^{n}E^n} \equiv
\sum_{m} \frac{1}{M} \vert m \rangle \langle m \vert_{\hat{M}} \otimes \omega_{A^{\prime}B^{n}E^{n}}%
^{m} .
\end{equation}
By using arguments similar to \eqref{eq:c4}--\eqref{eq:c10}, we find the following bound:
\begin{align}
\log_{2}M\leq I(\hat{M};A^{\prime}B^{n})_{\omega} - I(\hat{M};A^{\prime}E^{n})_{\omega}
+\left(  \varepsilon+\delta\right)  \log_2 M+g(\varepsilon)+g(\delta)~. 
\end{align}
Now by several applications of the chain rule for conditional mutual information,
we find that%
\begin{align}
I(\hat{M};A^{\prime}B^{n})_{\omega} - I(\hat{M};A^{\prime}E^{n})_{\omega}   
& = I(\hat{M} ; B^n \vert A^{\prime})_{\omega} - I (\hat{M}; E^n \vert A^{\prime})_{\omega} \\
& = \sum_{i=1}^{n} I(\hat{M}; B_i \vert B_1^{i-1} E_{i+1}^{n} A^{\prime})_{\omega} - I(\hat{M}; E_i \vert B_1^{i-1} E_{i+1}^{n} A^{\prime})_{\omega} \\
& \leq \sum_{i=1}^{n} \sup_{\N_{A \to A^{\prime \prime} A^{\prime \prime \prime}} } \left[I(A^{\prime\prime}; B_i \vert A^{\prime\prime\prime})_{\tau} - I(A^{\prime \prime} ; E_i \vert A^{\prime\prime\prime})_{\tau}\right] \\
& = n \sup_{\N_{A \to A^{\prime \prime} A^{\prime \prime \prime}} }  \left[ I(A^{\prime\prime}; B \vert A^{\prime\prime\prime} )_{\tau} - I (A^{\prime\prime}; E \vert A^{\prime\prime\prime})_{\tau} \right] ~,
\end{align}
where the first inequality follows because the action of
1)\ tensoring in the states $\rho_{ABE}^{\otimes i-1}$ and $\rho
_{ABE}^{\otimes n-i}$ to the $i$th copy of $\rho_{ABE}$, 2)\ tensoring in the state $\frac{1}{M}\sum_{m=1}%
^{M}|m\rangle\langle m|_{\hat{M}}$, 3)\ applying the encoding $\mathcal{E}%
_{A^{n}\rightarrow A^{\prime}}^{m}$ conditioned on the value $m$ in $\hat{M}$,
and 4)\ tracing over the systems $B_{i+1}^{n}$ and $E_{1}^{i-1}$ can be understood as a local
channel $\mathcal{N}_{A_{i}\rightarrow\hat{M}A^{\prime}B_1^{i-1}E_{i+1}^n}$ acting
on the $A_{i}$ system of $\rho_{A_{i}B_{i}E_{i}}$. Relabeling system $\hat{M}$ as $A''$ and systems $A^{\prime}B_1^{i-1}E_{i+1}^n$ as $A'''$, and defining
\begin{equation}
\tau_{A''A'''BE} \equiv \N_{A \to A^{\prime \prime} A^{\prime \prime \prime}}(\rho_{ABE}),
\end{equation}
we arrive at the inequality, following
from the fact that the supremum is taken over quantum channels $\N_{A \to A^{\prime \prime} A^{\prime \prime \prime}}$.

Putting everything together, we find the following bound on the number of bits that Alice can communicate securely:
\begin{equation}
\frac{1-\varepsilon-\delta}{n}\log_{2}M\leq  \sup_{\N_{A \to A^{\prime \prime} A^{\prime \prime \prime}} }  \left[ I(A^{\prime\prime}; B \vert A^{\prime\prime\prime} )_{\tau} - I (A^{\prime\prime}; E \vert A^{\prime\prime\prime})_{\tau} \right]+\frac
{g(\varepsilon)+g(\delta)}{n}.
\end{equation}
Taking the limit as $n\rightarrow\infty$ and then as $\varepsilon
,\delta\rightarrow0$ allows us to conclude that 
\begin{align}
\sup_{\N_{A \to A^{\prime \prime} A^{\prime \prime \prime}}}  \left[ I(A^{\prime\prime}; B \vert A^{\prime\prime\prime} )_{\tau} - I (A^{\prime\prime}; E \vert A^{\prime\prime\prime}) _{\tau} \right]
\end{align}
is an upper bound on the capacity of $\rho_{ABE}$ for this communication task.

\section{A proof of the direct coding theorem for the communication protocol in \cite{BO12}}

We now provide a proof of the direct coding theorem for the communication protocol described above, which follows from the coding scheme developed in Section \ref{sec:achievability-of-CQMI}. Let $M, K \in \mathbb{N}$. Alice has message variable $m\in \{1, \dots, M \}$ and a local key variable $k \in \{1, \dots, K \}$. Before communicating to Bob, Alice can apply any local operation to the $A^n$ systems. Suppose that Alice applies a quantum channel $\N_{A \to A'' A'''}$ to all $n$ copies of the state $\rho_{ABE}$. Then the overall state that describes all systems is
\begin{align}
\hat{\rho}^{\otimes n}_{A^{\prime \prime}A^{\prime \prime \prime}BE} = \N^{\otimes n}_{A \to A'' A'''} (\rho^{\otimes n}_{ABE})~.
\end{align}
Similar to the coding scheme developed in Section \ref{sec:achievability-of-CQMI}, if Alice wishes to send message $m$, then she picks $k$ uniformly at random from $k \in \{1, \dots, K \}$. For each pair $(m, k)$, the random code is selected in such a way that a vector $s$, of the form described in \eqref{eq:s-variable}, is chosen uniformly at random and associated with the pair $(m,k)$. So if Alice wishes to send the pair $(m,k)$, she applies the encoding unitary $U_{A^{\prime \prime n}}(s(m,k))$ to the state $\hat{\rho}^{\otimes n}_{A^{\prime \prime}A^{\prime \prime \prime}BE} $ and sends $A^{\prime \prime n}A^{\prime \prime \prime n}$ systems to Bob over the ideal quantum channel. Then we could structure a coding scheme similar to our achievability proof in Section \ref{sec:achievability-of-CQMI}, such that 
\begin{align}
\log_2 M K & \approx n I(A^{\prime \prime}; B A^{\prime \prime \prime}), \\
\log_2 K & \approx n I(A^{\prime \prime}; E A^{\prime \prime \prime})~. 
\end{align}
Then if 
\begin{align}
I(A^{\prime \prime}; B A^{\prime \prime \prime}) - I(A^{\prime \prime}; E A^{\prime \prime \prime})
\end{align}
is strictly positive, the coding scheme guarantees that this information difference is an achievable rate. Moreover, Alice can further improve the achievable rate of secure communication by optimizing over quantum channels $\N_{A \to A'' A'''}$. Therefore, the following rate is an achievable rate:
\begin{align}
\sup_{\N_{A \to A^{\prime \prime} A^{\prime \prime \prime}}}  \left[ I(A^{\prime\prime}; B \vert A^{\prime\prime\prime} ) - I (A^{\prime\prime}; E \vert A^{\prime\prime\prime}) \right]~. 
\end{align}
This concludes the achievability proof. 

\end{document}